\documentclass[]{aastex631}
\usepackage{amsmath}
\usepackage{amsthm}
\usepackage{amssymb}
\usepackage{mathrsfs}
\usepackage{extarrows}
\usepackage{color}
\usepackage{latexsym}
\usepackage{enumitem}
\usepackage{mathtools}
\usepackage{natbib}
\usepackage[utf8]{inputenc}
\DeclareUnicodeCharacter{02BC}{'}

\begin{document}

\title{Investigating an Erupting Metric-decimetric Radio 
Depression and its Physical Origin}

\author[0000-0001-7597-6620]{B. T. Wang}
\affiliation{School of Astronomy and Space Science, Nanjing University, Nanjing 210023, People's Republic of China; xincheng@nju.edu.cn}
\affiliation{Key Laboratory of Modern Astronomy and Astrophysics (Nanjing University), Ministry of Education, Nanjing 210093, People's Republic of China}
\email{wangbt@smail.nju.edu.cn}

\author[0000-0003-2837-7136]{X. Cheng}
\affiliation{School of Astronomy and Space Science, Nanjing University, Nanjing 210023, People's Republic of China; xincheng@nju.edu.cn}
\affiliation{Key Laboratory of Modern Astronomy and Astrophysics (Nanjing University), Ministry of Education, Nanjing 210093, People's Republic of China}
\email{wangbt@smail.nju.edu.cn}

\author[0000-0002-7597-7663]{J. Y. Yan}
\affiliation{State Key Laboratory of Space Weather, National Space Science Center, Chinese Academy of Sciences, Beijing 100190, Peopleʼs Republic of China; yanjingye@nssc.ac.cn}
\affiliation{Radio Science and Technology Center ($\pi$ Center), Chengdu 610041, Peopleʼs Republic of China}
\email{wangbt@smail.nju.edu.cn}

\author[0000-0002-6550-1522]{C. Xing}
\affiliation{School of Astronomy and Space Science, Nanjing University, Nanjing 210023, People's Republic of China; xincheng@nju.edu.cn}
\affiliation{Key Laboratory of Modern Astronomy and Astrophysics (Nanjing University), Ministry of Education, Nanjing 210093, People's Republic of China}
\email{wangbt@smail.nju.edu.cn}

\author{W. T. Fu}
\affiliation{School of Astronomy and Space Science, Nanjing University, Nanjing 210023, People's Republic of China; xincheng@nju.edu.cn}
\affiliation{Key Laboratory of Modern Astronomy and Astrophysics (Nanjing University), Ministry of Education, Nanjing 210093, People's Republic of China}
\email{wangbt@smail.nju.edu.cn}

\author{L. Wu}
\affiliation{State Key Laboratory of Space Weather, National Space Science Center, Chinese Academy of Sciences, Beijing 100190, Peopleʼs Republic of China; yanjingye@nssc.ac.cn}
\affiliation{Radio Science and Technology Center ($\pi$ Center), Chengdu 610041, Peopleʼs Republic of China}
\email{wangbt@smail.nju.edu.cn}

\author{L. Deng}
\affiliation{State Key Laboratory of Space Weather, National Space Science Center, Chinese Academy of Sciences, Beijing 100190, Peopleʼs Republic of China; yanjingye@nssc.ac.cn}
\affiliation{Radio Science and Technology Center ($\pi$ Center), Chengdu 610041, Peopleʼs Republic of China}
\email{wangbt@smail.nju.edu.cn}

\author{A. L. Lan}
\affiliation{State Key Laboratory of Space Weather, National Space Science Center, Chinese Academy of Sciences, Beijing 100190, Peopleʼs Republic of China; yanjingye@nssc.ac.cn}
\affiliation{Radio Science and Technology Center ($\pi$ Center), Chengdu 610041, Peopleʼs Republic of China}
\email{wangbt@smail.nju.edu.cn}

\author{Y. Chen}
\affiliation{State Key Laboratory of Space Weather, National Space Science Center, Chinese Academy of Sciences, Beijing 100190, Peopleʼs Republic of China; yanjingye@nssc.ac.cn}
\affiliation{Radio Science and Technology Center ($\pi$ Center), Chengdu 610041, Peopleʼs Republic of China}
\email{wangbt@smail.nju.edu.cn}

\author{C. Wang}
\affiliation{State Key Laboratory of Space Weather, National Space Science Center, Chinese Academy of Sciences, Beijing 100190, Peopleʼs Republic of China; yanjingye@nssc.ac.cn}
\affiliation{Radio Science and Technology Center ($\pi$ Center), Chengdu 610041, Peopleʼs Republic of China}
\email{wangbt@smail.nju.edu.cn}

\author[0000-0002-4978-4972]{M. D. Ding}
\affiliation{School of Astronomy and Space Science, Nanjing University, Nanjing 210023, People's Republic of China; xincheng@nju.edu.cn}
\affiliation{Key Laboratory of Modern Astronomy and Astrophysics (Nanjing University), Ministry of Education, Nanjing 210093, People's Republic of China}
\email{wangbt@smail.nju.edu.cn}

\begin{abstract}

We present direct metric-decimetric radio imaging observations of a fascinating quiescent filament eruption on 2024 March 17 using data from the DAocheng Radio Telescope (DART), with a combination of the Solar Dynamics Observatory and the Chinese H$\alpha$ Solar Explorer. At the radio band, even though the filament is difficult to identify in its early phase, it rapidly became distinct and formed a continuous loop-like dark structure during the eruption, i.e., so-called radio depression. Compared with the fragmentation of the erupting filament observed at the H$\alpha$ and EUV bands, the radio depression appeared more coherently.
Based on synthetic radio images from a three-dimensional magnetohydrodynamics (MHD) simulation of a flux-rope-filament eruption, it is suggested that the radio depression originates from the absorption of cold and dense materials within the erupting flux rope to the background emission. The absorption seems to be stronger than that at the H$\alpha$ and EUV bands, thus leading to their apparent discrepancies. Moreover, the radio depression is also found to occupy the lower part but not the whole body of the flux rope. 

\end{abstract}


\section{Introduction} \label{sec:1}

Solar filaments are cold and dense plasma suspended in the hot and tenuous corona. They typically appear as absorption features against a bright background as seen on the solar disk. Filaments alternatively appear as bright structures, known as prominences, when observed above the solar limb \citep[for a review, see][]{Parenti2014}. Numerous studies support the idea that the filament is typically a collection of cool materials at the dips of sheared arcades or a magnetic flux rope (MFR), the latter of which is a twisted flux rope with helical field lines wrapped around a common axis \citep[e.g.,][]{Aulanier1998, Guo2010, Xia2012, Jiang2014}.

Filaments may erupt suddenly and give rise to coronal mass ejections (CMEs), thus being of great significance for predicting space weather \citep{Schmieder2002, Gopalswamy2003, Parenti2014}. White-light coronagraph observations show that CMEs often display a three-part component structure including the bright core, dark cavity, and bright front, the former two of which are believed to correspond to the erupting filament and the whole flux rope structure, respectively \citep{Illing1985, Webb1987, Vourlidas2013}. Due to the great visibility of filaments at multi-wavelengths such as extreme UV (EUV), H$\alpha$, and white light, their kinematics have also been extensively studied for the sake of uncovering the initiation and early evolution of CMEs \citep[e.g.,][]{Heinzel2001, Cheng2013, McCauley2015, Song2019, Cheng2020, Song2022,  Wang2023}. 

Despite filaments having been observed for decades, radio imaging observations of filament eruptions, particularly at the metric and decimetric wavebands, remain rather rare \citep[for a brief review see][]{Alissandrakis2020}. Using Nan\c{c}ay Radioheliograph (NRH) data, \citet{Marque2002} reported a depression structure appearing at 410.5 MHz and found that it showed a significant discrepancy with the filament as seen at the 304 \AA. Subsequently, \citet{Marque2004} performed a systematic study on the nature of filament-related radio depressions and proposed that the depressions corresponded to a dark cavity surrounding the filament, where the plasma density was relatively low. On the other hand, contradictory results were also reported. It was found that radio depressions at the centimeter and millimeter bands were of a similar scale to H$\alpha$ filaments, indicating that they were both caused by the absorption of cold and dense plasma \citep{Kundu1972,Raoult1979,Alissandrakis2013,Alissandrakis2020}.

In this paper, we report metric-decimetric radio imaging observations of a quiescent filament eruption using the DAocheng Radio Telescope \citep[DART, formerly known as the DSRT;][]{Yan2023}. For simplicity, the filament-related radio brightness depression, i.e., \textit{Radio Depression}, is adopted in the following. Our observations show that, unlike the fragmentation of the filament observed at the H$\alpha$ and EUV bands, the radio depression appears more coherently, even becoming darker during the eruption. Based on the comparison with a magnetohydrodynamics (MHD) simulation of an MFR eruption, we suggest that the radio depression originates from the absorption of cold plasma and can serve as a tracer of the erupting MFR. The data and methods are introduced in Section \ref{sec:2}. In Section \ref{sec:3}, we present the results, which are followed by a brief summary and discussion in Section \ref{sec:4}.

\section{Instruments and Methods} \label{sec:2}

The metric-decimetric radio imaging data are from the DART that operates at frequencies between 150 MHz and 450 MHz. Depending on the frequency, the field of view (FOV) and the spatial resolution range from $7^{\circ}$ and $1^{\prime}.7$ (at 450 MHz) to $20^{\circ}$ and $5^{\prime}.2$ (at 150 MHz), respectively. The time cadence of radio images we used is 10.5 s. DART is a circular array of 1 km diameter with 313 element antennas, each with an aperture of 6 m. Cross correlation between each pair of antennas represents a spatial frequency component in the u-v domain. These measurements, referred to as visibilities, sample Fourier harmonics of the brightness distribution within the FOV. By acquiring a sufficient number of visibilities (approximately 100,000 in dual polarizations for DART), a solar image can be reconstructed using Fourier inversion. At present, although the calibration of absolute brightness temperature is still under way, the normalized brightness temperature is reliable for further analysis after a comparison with the data from the NRH and Murchison Widefield Array (MWA) (see Section \ref{subsec:3.1}). Here, the radio image 20 mins before the filament eruption was used to do normalization. The EUV data are from the Atmospheric Imaging Assembly \citep[AIA;][]{Lemen2012} on board the Solar Dynamics Observatory (SDO). The spatial resolution and time cadence are about $1^{\prime\prime}.5$ and 12 s, respectively. To trace the evolution of the filament, we also took advantage of the data from the H$\alpha$ Imaging Spectrograph (HIS) of the Chinese H$\alpha$ Solar Explorer \citep[CHASE; ][]{Li2022}, which provides full-disk H$\alpha$ spectroscopic images through the scanning mode at the H$\alpha$ band of 6559.7-6565.9 \AA\ with the spectral sampling of 0.0484 \AA\ and the spatial resolution of $\sim$ $2^{\prime\prime}$ \citep{Qiu2022}. The Large Angle and Spectrometric Coronagraph \citep[LASCO; ][]{Bruecnker1995} on board the Solar and Heliospheric Observatory (SOHO) provides the white-light images of associated CME.

\begin{figure}[ht!]
\plotone{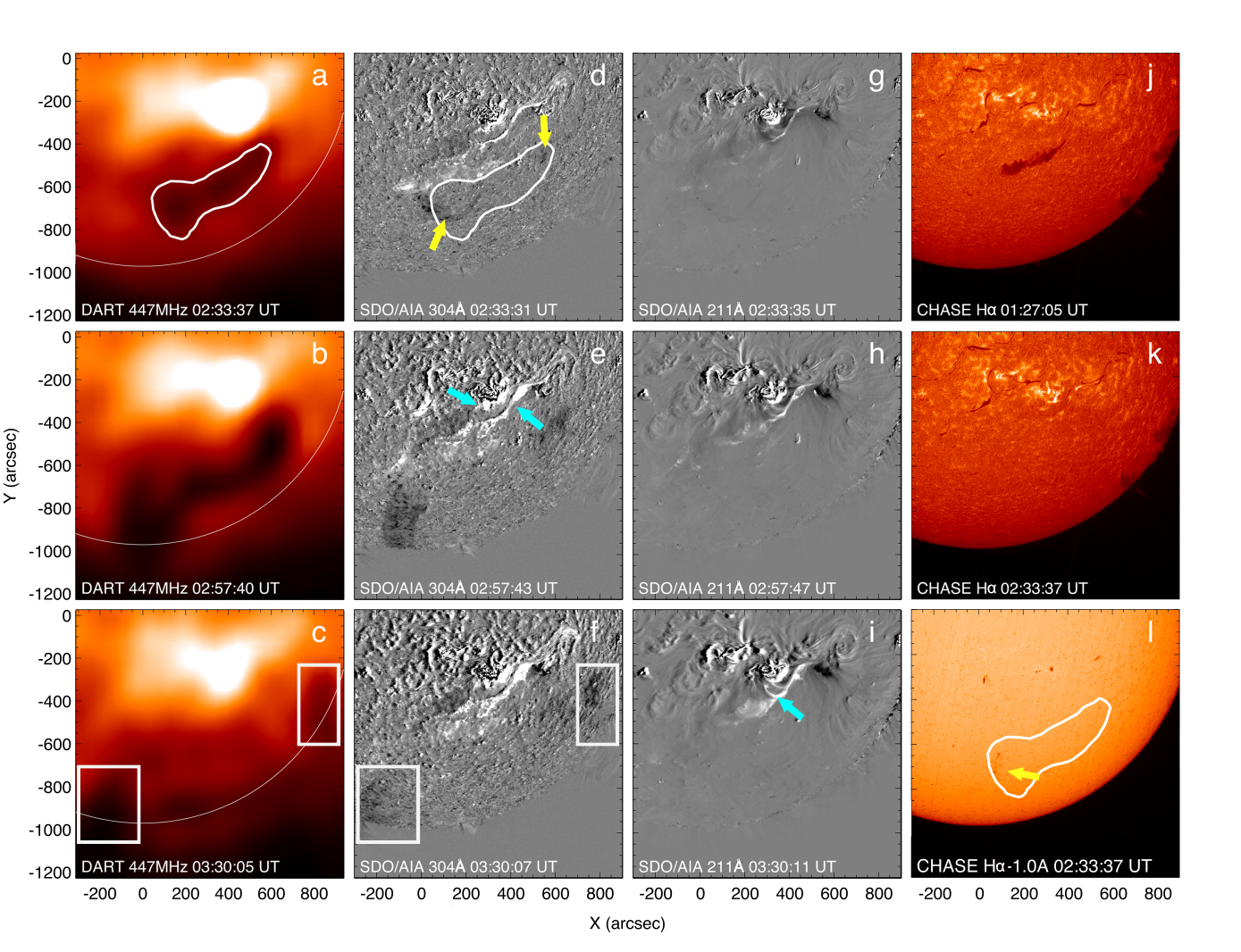}
\caption{(a)-(c) DART 447 MHz images showing the evolution of the radio depression with the solar limb indicated by the partial circle in white. (d)-(f) and (g)-(i) Same as panels (a)-(c) but for the AIA 304 \AA\ and AIA 211 \AA\ base difference images, respectively. (j) CHASE H$\alpha$ line center image showing the pre-eruptive filament. (k) and (l) CHASE H$\alpha$ line center and blue wing 1.0 \AA\ images evidencing the invisibility of the filament during the eruption. The contours in white outline the radio depression as shown in panel (a). The arrows in yellow indicate the EUV and H$\alpha$ filament during the early stage. The arrows in cyan mark the flare ribbons and post flare loops. The white boxes mark the visible part of the radio depression and EUV filament. The evolution of the erupting filament at the AIA 211 \AA\ and 304 \AA\ between 02:00 UT and 04:00 UT is shown in the animation. The real-time duration of the animation is 12 s. \\
(An animation related to this figure is available.)
\label{fig:1}}
\end{figure}

To determine the nature of the radio depression, we also synthesize the radio and EUV images of a simulated filament eruption. Here, we only consider the Bremsstrahlung emission because it dominates the emissivity of filaments at the metric and decimetric bands \citep{Alexander2020}. This is justified by the following two facts: (1) the filament-associated flare is very weak and the population of non-thermal electrons is small; the induced coherent emissions may be insignificant; (2) the filament eruption is from the region with a relatively weak magnetic field, and thus the gyromagnetic emission can be further neglected \citep{Shibasaki2011}. 
We follow \citet{Dulk1985} and \citet{Draine2011} to calculate the radio radiation as follows: the bremsstrahlung emissivity coefficient $ j_{\nu}^{ff}$ and absorption coefficient $ \kappa_{\nu}^{ff}$ are expressed as:

\begin{equation} \label{eq:1}
j_{\nu}^{ff} \approx 9.78 \times 10^{-3}\frac{n_{e}}{T^{1/2}}e^{-h\nu/kT}\sum_{i}Z_{i}^2n_{i} \\
\times
\begin{dcases}
18.2 + 1.5 lnT - ln\nu& \text{$(T \leq 2\times10^5 K)$} \\
24.5 + lnT - ln\nu& \text{$(T \geq 2\times10^5 K)$}
\end{dcases}
\end{equation}

\begin{equation} \label{eq:2}
\kappa_{\nu}^{ff} \approx  9.78 \times 10^{-3}\frac{n_{e}}{\nu^{2}T^{3/2}}\sum_{i}Z_{i}^2n_{i} \\
\times
\begin{dcases}
18.2 + 1.5 lnT - ln\nu& \text{$(T \leq 2\times10^5 K)$} \\
24.5 + lnT - ln\nu& \text{$(T \geq 2\times10^5 K)$}
\end{dcases}
\end{equation}

where, $\nu$ is the frequency of radio emission, $Z_{i}$ represents the charge of an ion, $n_{i}$ and $n_{e}$ are the number densities of ions with charges of $Z_{i}e$ and electrons, respectively. $T$ denotes the plasma temperature. 

Based on the formula above, we use the Radiation Synthesis Tools \citep{Fuwentai} to synthesize the optically thick images of the simulated filament at the 171 \AA\ and 300 MHz in multiple perspectives. The optically thick radiation synthesis at the EUV passbands is also achieved by RST. More details about the radiation synthesis methods can be found in \citet{Fuwentai}.

\section{Results} \label{sec:3}
\begin{figure}[ht!]
\centering
\resizebox{1.0\textwidth}{!}{\plotone{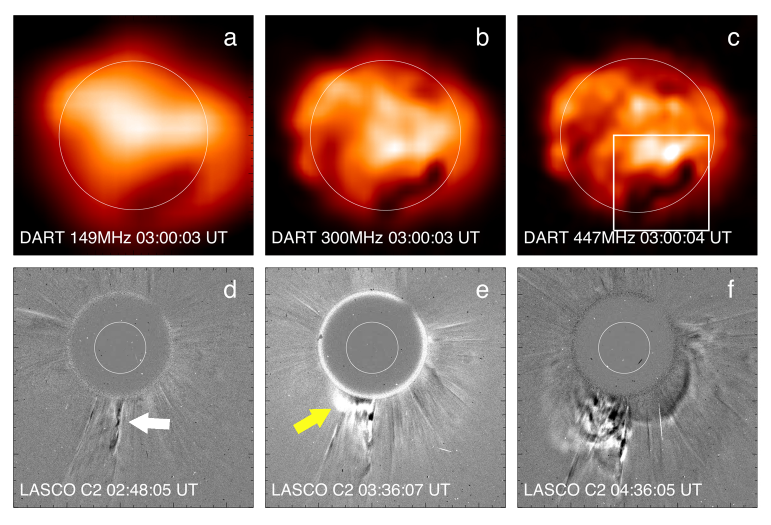}}
\caption{(a)-(c) DART radio images of the erupting filament reconstructed at the frequencies of 149 MHz, 300 MHz, and 447 MHz, respectively. The white circle represents the optical solar limb. 
(d)-(f) LASCO C2 images showing the appearance of two CMEs. The white arrow marks the first narrow CME, which is unrelated to the filament we studied. The yellow arrow mark the first appearance of the partial halo CME caused by the filament eruption. The evolution of the radio depression at three frequencies (149 MHz, 300 MHz and 447 MHz) between 02:00 UT and 04:00 UT is shown in the animation. The real-time duration of the animation is 13 s. \\
(An animation related to this figure is available.)
\label{fig:2}}
\end{figure}

The quiescent filament of interest erupted on 2024 March 17th, accompanied by a C2.6 class flare\footnote{For flare information, please refer to: \url{https://www.lmsal.com/solarsoft/latest_events_archive/events_summary/2024/03/17/gev_20240317_0205/index.html}}.
After a gradual increase, the GOES 1-8 \AA\ soft X-ray flux peaked at 02:36 UT. The flare ribbons and post flare loops (cyan arrows in Figure \ref{fig:1}) can be clearly identified at the EUV passbands (for more details, please see the animation accompanying Figure \ref{fig:1}). Interestingly, the DART observed the entire eruption process, showing that the filament was closely associated with the region of pronounced brightness depression in the metric-decimetric images, even though it was less visible at the beginning (see Figure \ref{fig:2} and the animation accompanying Figure \ref{fig:2}). In contrast, the filament was clearly visible in the CHASE H$\alpha$ line center images prior to the eruption (Figure \ref{fig:1} (j)). Around 01:20 UT, it began to rise slowly, as clearly identified in the AIA EUV images (see the animation accompanying Figure \ref{fig:1}). After a period of acceleration ($\sim$ 1 hour), the filament gradually became invisible in the H$\alpha$ line center images but appeared in the blue wing images, indicating a Doppler shift caused by the outward eruption of the cool materials. The propagation of the filament eruption can be well traced at both the EUV and radio passbands until exceeding the solar limb. The eruption finally resulted in a faint partial halo CME as observed by the LASCO C2 coronagraph after 03:36 UT (marked with yellow arrows in Figure \ref{fig:2} (e)). Note that, another narrow CME appeared before 02:30 UT (marked with the white arrow in Figure \ref{fig:2} (d)) is not related to the eruption we studied after a careful inspection. This filament eruption was also studied by \citet{Hou2025} for the sake of searching for sun-as-a-star signatures of filament eruptions at the radio band.

\subsection{Radio Imaging of the Filament Eruption \label{subsec:3.1}}
\begin{figure}[ht!]
\centering
\resizebox{1.0\textwidth}{!}{\plotone{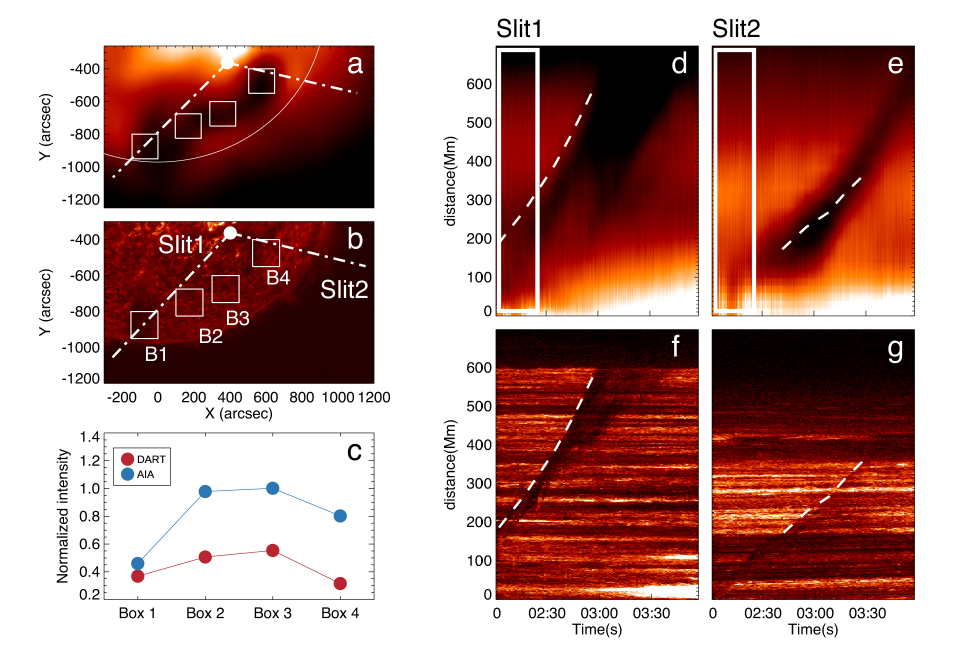}}
\caption{(a)-(b) DART 447 MHz and AIA 304 \AA\ images at 02:57 UT, respectively. (c) Normalized intensity in Box 1-4 (white boxes in panel (a) and (b)) at 02:57 UT. (d) and (e) Distance-time plots along Slit 1 and Slit 2 (white dotted-dashed line in panel (a) and (b)) representing the evolution of the radio depression during the eruption. (f)-(g) same as panels (d)-(e) but for the AIA 304 \AA\ passband. The white dashed line shows the EUV filament front. 
\label{fig:3}}
\end{figure}

At the metric and decimetric wavebands, the radio depression first became visible at $\sim$ 02:30 UT, appearing as a distinct dark structure and becoming darker during its propagation in the following 1 hour.
Interestingly, the radio depression maintained a continuous, loop-like structure at different frequencies for a long time (Figure \ref{fig:2} (a)-(c)). We compared the spatial extension of the filament at multi-passbands (see Figure \ref{fig:1} (a), (d), (k), and (l)). Unlike the continuous depression at the radio band, a large portion of the filament is invisible at the EUV and H$\alpha$ bands during the eruption, with only a small part of it visible above the two footpoints.
The propagation of the filament along two slits is displayed in the Distance-time plots as shown in Figure \ref{fig:3}. The early rise of the radio depression before $\sim$ 02:30 UT (white boxes in Figure \ref{fig:3}) was difficult to discern due to its ambiguity, whereas the EUV filament was clearly identifiable. 
Shortly after, the radio depression underwent an apparent expansion both in angular width and cross section between 02:30 UT and 03:00 UT (see Figure \ref{fig:1} (a)-(c) and Figure 3) but with the former being more obvious than the latter. Meanwhile, the EUV filament also experienced a similar expansion. After 03:30 UT, the middle part of the radio depression gradually became invisible, while the remaining portion was approximately co-spatial with the EUV filament (indicated by the white boxes in Figure \ref{fig:1} (c) and (f)). Moreover, the front of the EUV filament is found to be aligned with that of the radio depression along Slit 1 but slightly lagged behind that along Slit 2. To further address the relevance between the radio depression and the EUV filament, we calculate the normalized intensity with DART and AIA 304 \AA\ data for four boxes along the radio depression (Figure \ref{fig:3} (c)). A consistent variation tendency can be clearly observed, the normalized DART intensity is smaller for the boxes that include obviously visible EUV filament materials (B1 and B4).
It indicates that they are closely related to each other but may be distinct in formation.

\begin{figure}[ht!]
\centering
\resizebox{1.0\textwidth}{!}{\plotone{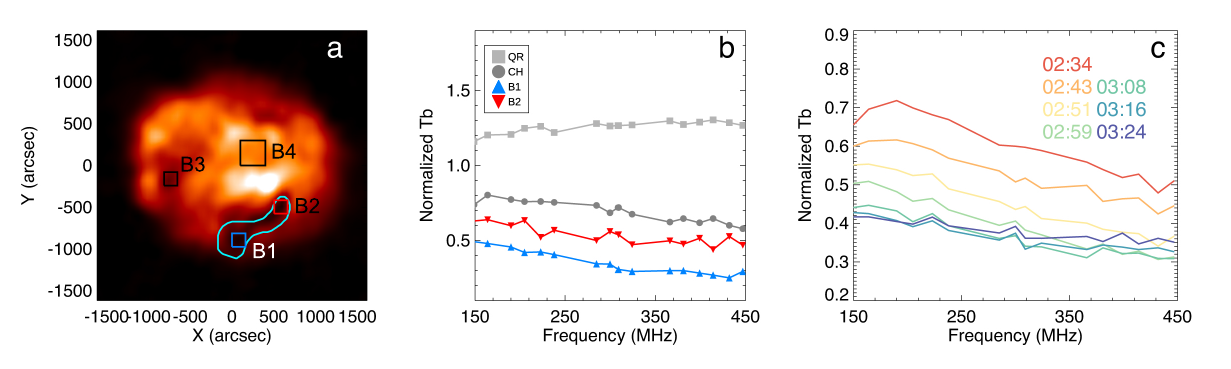}}
\caption{(a) DART 447 MHz images at 02:57 UT. The cyan contour outlines the radio depression. B1 and B2 are two boxes at the radio depression. B3 and B4 are boxes in the coronal hole (CH) and quiescent region (QR), respectively. (b) Normalized brightness temperature (Tb) as a function of frequency for B1-B4. (c) Variation of the normalized brightness temperature over time at different frequencies. 
\label{fig:4}}
\end{figure}

A comparison of normalized brightness temperature for the radio depression and nearby undepressed regions is shown in Figure \ref{fig:4} (b). It is noted that the normalized brightness temperature slightly decreases with frequency for both the depressions (B1 and B2) and the coronal hole (B3), while it almost keeps a constant around one for the quiescent region (B4). This is consistent with the results from the NRH and MWA, where coronal holes appear as darker structures at higher frequencies \citep{Mercier2012, Rahman2019}.
As the filament erupted, the normalized brightness temperature further decreased at all frequencies between 02:30 UT and 03:00 UT (Figure \ref{fig:3} (d) and (e)). It could be a consequence of the decrease of the foreground radiation as the filament rose (see Appendix \ref{sec:apd} for more details). 

\subsection{Modeling the Radio Depression\label{subsec:3.2}}

To determine the nature of the radio depression, we synthesize the radio images, taking advantage of MHD simulation data of an MFR eruption involving a filament \citep{Xingchen}. The simulation, an updated version of \citet{Xing2024a}, is performed by the code MPI-AMRVAC \citep{Xia2018}. It solves the full-MHD equations and takes the resistivity, viscosity, gravity, thermal conduction, radiative cooling, and background heating into account. The highest spatial resolution is about 300 km in the horizontal direction and 150 km in the vertical direction, being comparable to that of the AIA images. The simulation starts with a potential bipolar field and a hydrostatic atmosphere ranging from the chromosphere to the corona. After a thermal relaxation process, the initial field is firstly driven to a highly sheared state with imposed line-tied bidirectional shearing flows at the two sides of the polarity inversion line (PIL). Later, driven by the converging flows towards the PIL, a pre-eruptive MFR is formed by magnetic reconnection. A filament is also naturally formed during the MFR build-up phase. Please refer to \citet{Xingchen} for more details on the formation of the MFR and associated filament. 

\begin{figure}[ht!]
\centering
\resizebox{1.0\textwidth}{!}{\plotone{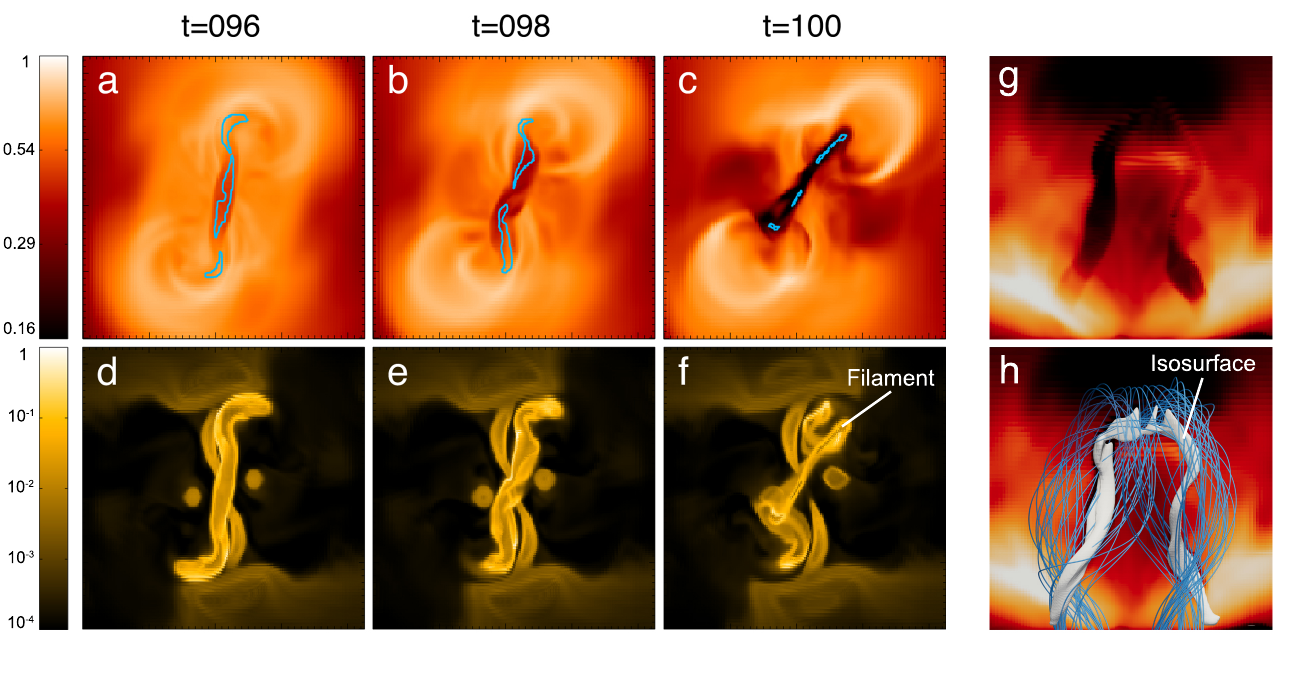}}
\caption{Radio and EUV synthetic images based on the MHD simulation of a flux-rope-filament eruption. (a)-(c) Synthetic 300 MHz radio images at t=96, 98, and 100. (d)-(f) Same as panels (a)-(c) but for the AIA 171 \AA\ passband. Cyan contours in panels (a)-(c) outline the EUV filaments in panels (d)-(f). (g)-(h) Synthetic 300 MHz radio images at t=100 from a side view. A white isosurface, which corresponds to plasma density approximately five times greater and plasma temperature about one order of magnitude lower than those of the quiet corona, along with magnetic field lines of the erupting flux rope, is overplotted in panel (h).
\label{fig:5}}
\end{figure}

Figure \ref{fig:5} (a)-(f) show synthetic images of the modeled flux-rope-filament system at the frequency of 300 MHz and at the AIA 171 \AA\ passband, which are calculated by the Radiation Synthesis Tools. The EUV filament was comparable to the radio depression at the beginning of the eruption ($t = 96$) but became fragmented and much smaller than the radio depression during the eruption (e.g., $t = 100$). Furthermore, the radio depression became darker and longer at 300 MHz during the eruption, which are attributed to the foreground reduction and expansion, respectively. Such characteristics in the visibility of the erupting filament between the EUV and metric-decimetric bands somehow resemble what are disclosed in observations. 
Note that the modeled radio depression also shows some different characteristics from observations. The modeled one is relatively slender and does not display a remarkable expansion in width. This could be attributed to two factors. On the one hand, part of the modeled filament is heated by magnetic reconnection, which is much stronger than that for the case we observed, and thus does not appear as dark structures at the radio band. On the other hand, due to the limitation of the simulation in box size, only the early stage of the filament eruption is derived ($t = 100$ corresponds to approximately 10 mins after the onset of the eruption), the following expansion hence could be missed.
We further compare the spatial relationship between the MFR and radio depression as shown in Figure \ref{fig:5} (h). It is found that the radio depression is roughly co-spatial with the lower part of the MFR below the main axis, not occupying the entire MFR structure. 

\section{Summary and Discussion} \label{sec:4}
In this paper, we present a detailed study on the metric-decimetric imaging observations of a quiescent filament eruption. 
The radio depression was extremely faint during the initiation phase of the eruption but rapidly became darker and more distinct after the eruption of the H$\alpha$ filament, manifesting as a continuous loop-like structure. A comparison of multi-passband images shows that the radio depression seems to be larger than the EUV and H$\alpha$ filament in space. In particular, the former always kept a coherence while the latter presented a fragmentation during the eruption. Moreover, the radio depression also exhibited stronger darkening at higher frequencies, with normalized brightness temperatures decreasing over time during the eruption.
To further study the physical origin of the radio depression, we synthesize radio and EUV images based on the MHD simulation of an MFR/filament eruption, which show some similarities with the observations. It is suggested that the radio depression originates from the absorption of the cool and dense materials within the MFR. The absorption seems to be stronger at the metric-decimetric bands than that at the EUV and H$\alpha$ bands, thus resulting in a disparity in visibility.

The nature of the radio depression at the metric and decimetric bands has been mentioned previously in \citet{Marque2002}, where it was interpreted as the cavity with a lower plasma density than the background. In our study, there is a strong correlation between the evolution of the EUV filament and that of the radio depression during the eruption (Section \ref{subsec:3.1}), which indicates a close corelation existing between the radio depression and the cool plasma.
The simulated radio depression is found to be almost co-spatial with the high-density and low-temperature structure (as shown by the isosurface in Figure \ref{fig:5} (h)), further suggesting that it is the absorption from the cold and dense plasma leads to the decrease of the metric and decimetric radio emissions, rather than the reduction of the plasma density within the cavity, in agreement with the interpretation of \citet{Hou2025}. This is also supported by a simple estimation using Equation \ref{eq:2}. With an optical thickness $\tau = \kappa ds = 1$ and typical parameters of filaments ($n_{e} = 10^{10} cm^{-3}, T = 10^4 K$) and corona ($n_{e} = 10^{8} cm^{-3}, T = 10^6 K$), $ds$ are estimated to be $\sim$ 50 m and $\sim$ 400 Mm, respectively. It means that the radio radiation can be strongly absorbed by cool filament materials. During the expansion stage, the plasma density within the MFR decreases, weakening its absorption of the solar background radiation. However, the foreground radiation above the filament decreases more rapidly at the beginning (see Appendix \ref{sec:apd} for details), leading to the darkening of the radio depression. As the expansion progresses, the foreground radiation becomes increasingly weak, and the plasma density within the radio depression continues to decrease, ultimately make it invisible. This is consistent with the result that no clear bright core is observed in the CME.
In contrast, the EUV filament appears to be fragmented early in the eruption (shown in Figure \ref{fig:5} (f)), which could be due to the fact that only sufficiently cool and dense materials can appear to be dark at the EUV bands.

On the other hand, \citet{Marque2004} proposed that the radio depression corresponds to the entire MFR surrounding the filament. Our study, however, shows that the radio depression only occupies the lower part of the MFR. 
In spite of this, given the coherency of the radio depression and the fragmentation of the EUV and H$\alpha$ filament, we suggest that the radio depression can still serve to track the evolution of the MFR when studying the early dynamics of CMEs.

\begin{acknowledgments}
We would like to thank Shaheda Begum Shaik in George Mason University for the valuable discussions. This paper utilizes data from the DART of the Meridian Project led by National Space Science Center and the CHASE mission supported by the China National Space Administration. This study is supported by the NSFC under grant 12403066, the Jiangsu NSF under grant BK20241187, the Fundamental Research Funds for the Central Universities under grant 2024300348, the Postdoctoral Fellowship Program of CPSF under grant GZC20240693, and the Jiangsu Funding Program for Excellent Postdoctoral Talent.
\end{acknowledgments}

\appendix
\section{Enhancement of the radio depression}\label{sec:apd}
In Section \ref{sec:3}, the radio depression was found to be hardly discernible before and in the early phase of the eruption. Then it quickly darkened, accompanied by its rise and expansion. 
\begin{figure}[ht!]
\centering
\renewcommand{\thefigure}{A1}
\resizebox{0.6\textwidth}{!}{\plotone{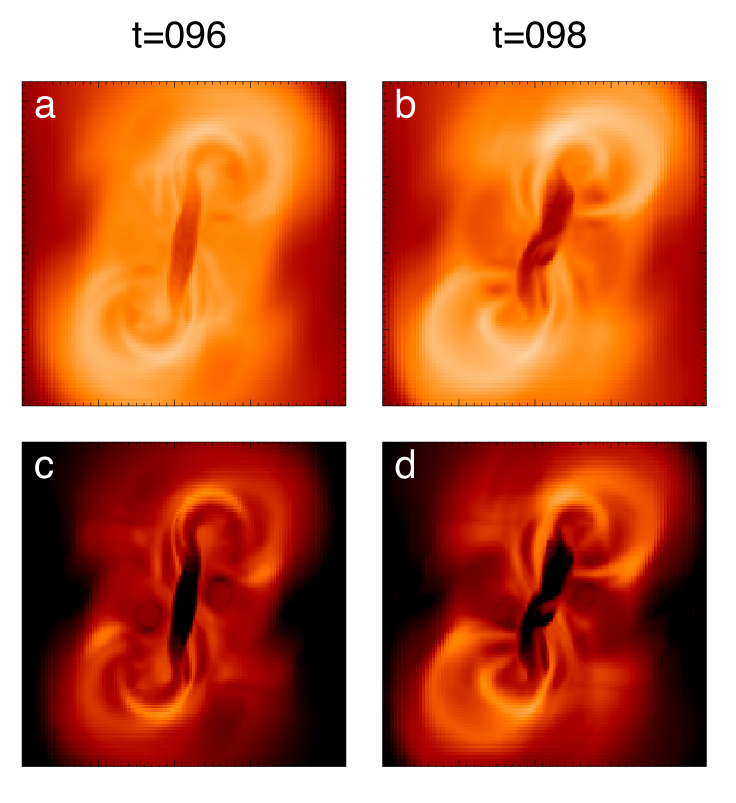}}
\caption{Comparison of synthetic images with and without foreground radiation. (a)-(b) The synthetic 300 MHz radio images at t=96, 98. (c)-(d) Same as (a)-(b) but excluding the foreground radiation. \label{fig:A1}}
\end{figure}
To investigate the nature of this darkening, we synthesize the radio images by removing the foreground radiation above the filament (Figure \ref{fig:A1} (c) and (d)). In this case, the radio depression remains distinct throughout the eruption, with no further darkening over time. This suggests that radiation from the foreground plays a substantial role at metric and decimetric bands, potentially covering the radio depressions associated with filaments when they are located in the low corona.

\bibliography{sample631}{}
\bibliographystyle{aasjournal}

\end{document}